\documentclass[showpacs,showkeys,preprintnumbers,amsmath,amssymb,latexsym]{revtex4}

\usepackage{dcolumn}

\def\fl{}
\def\rmd{{\rm d}}

\catcode`@=11  
\def\meqalign#1{\null\,\vcenter{\openup\jot\m@th
  \ialign{\strut\hfil$\displaystyle{##}$&&$\displaystyle{{}##}$\hfil
     \crcr#1\crcr}}\,}
\catcode`@=12  

\def\pmb#1{\setbox0=\hbox{$#1$}%
  \kern-.025em\copy0\kern-\wd0
  \kern.05em\copy0\kern-\wd0
  \kern-.025em\raise.0433em\box0}
\def\bfrho{{\pmb{\rho}}}  \def\bftheta{{\pmb{\theta}}}
\def\bfnabla{{\pmb{\nabla}}}
\def\mathbfcalJ{{\pmb{\mathcal{J}}}}

\begin{document}

\preprint{}

\title{Massless field perturbations and gravitomagnetism \\
in the Kerr-Taub-NUT spacetime}

\author{D. Bini}
  \email{binid@icra.it}
 \affiliation{Istituto per le Applicazioni del Calcolo ``M. Picone'', CNR I-00161 Rome, Italy and\\ 
International Center for Relativistic Astrophysics - I.C.R.A.\\
University of Rome ``La Sapienza'', I-00185 Rome, Italy}
\author{C. Cherubini}
  \email{cherubini@icra.it}
 \affiliation{Dipartimento di Fisica ``E.R. Caianiello'', Universit\`a di Salerno, I-84081, Italy,\\
International Center for Relativistic Astrophysics - I.C.R.A.\\
University of Rome ``La Sapienza'', I-00185 Rome, Italy\\
Institute of Cosmology and Gravitation, University of Portsmouth, Portsmouth PO1 2EG, England, UK
}
\author{R.T. Jantzen}
  \email{robert.jantzen@villanova.edu}
 \affiliation{Department of Mathematical Sciences, Villanova University,
  Villanova, PA 19085, USA and \\
International Center for Relativistic Astrophysics - I.C.R.A.\\
University of Rome ``La Sapienza'', I-00185 Rome, Italy}
\author{B. Mashhoon}
 \email{MashhoonB@missouri.edu}
\affiliation{Department of Physics and Astronomy,
University of Missouri-Columbia, Columbia,
Missouri 65211, USA}

\date{\today}

\begin{abstract}
A single master equation is given describing spin $s\le2$ test fields
that are 
gauge- and tetrad-invariant perturbations of the Kerr-Taub-NUT spacetime representing a source with mass $M$, gravitomagnetic monopole moment $-\ell$ and gravitomagnetic dipole moment (angular momentum) per unit mass $a$.
This equation can be separated into its radial and angular parts. The behavior of the radial functions at infinity and near the horizon is studied and used to examine the influence of
$\ell$ on the phenomenon of superradiance, while the angular equation leads to spin-weighted spheroidal harmonic solutions generalizing those of the Kerr spacetime.
Finally the coupling between the spin of the perturbing field and the gravitomagnetic monopole moment is discussed.
\end{abstract}


\keywords{Kerr-Taub-NUT, perturbations, master equation
          \\[12pt]
          Submitted to Phys.\ Rev.\ D December 6, 2002;
          revised and resubmitted January 21, 2003
}

\maketitle

\section{Introduction}

The Kerr-Taub-NUT (KTN) spacetime \cite{ES},
first discovered by Demia\'nski and Newman \cite{DemNew} and later studied by many others (see \cite{Miller} for references and a global analysis),
describes a stationary axisymmetric object with gravitomagnetic monopole and dipole moments associated with nonzero values of the NUT and Kerr parameters $\ell$ and $a$ respectively, and as such is a useful model for exploring gravitomagnetism. The well-known Dirac quantization of the magnetic monopole corresponds in the gravitomagnetic case to a condition first found by Misner \cite{Misner,MisTau} for the nonrotating special case of the Taub-NUT spacetime \cite{NUT} (zero Kerr parameter) that is consistent with forcing periodicity in the time coordinate, making the gravitomagnetic monopole Taub-NUT spacetime unphysical {\it under normal circumstances} by having closed timelike lines, but nonetheless interesting as a laboratory for probing other consequences of gravitomagnetic monopoles. As discussed by Miller \cite{Miller}, this same global structure is also consistent with the more general Kerr-Taub-NUT spacetime, which is the rotating version of this simpler spacetime, its rotation being associated with the nonzero gravitomagnetic dipole moment. 

The Kerr-Taub-NUT spacetime and its special cases all belong to the larger class of stationary axisymmetric type D vacuum solutions of the Einstein equations found by Carter \cite{Carter} for which the Hamilton-Jacobi equation for geodesics is separable. The stability of these spacetimes is probed by studying their perturbations by fields of various spin. Perturbations by massless fields have been investigated in the two limiting cases of the Kerr spacetime (pure gravitomagnetic dipole) \cite{bcjr} and the Taub-NUT spacetime (pure gravitomagnetic monopole) \cite{TeukTN} in a single unifying approach making use of the de Rham Laplacian which allows all fields of spin 0 through 2 to be considered together. 
In this article, the same analysis is applied to the Kerr-Taub-NUT spacetime.
Motivated by scattering off virtual black holes, Prestidge \cite{Prestidge} has performed this analysis for all spins except 3/2 for  the vacuum C-metric, another vacuum type D metric representing a pair of uniformly accelerated gravitoelectric monopoles.  

This approach to perturbations builds on the pioneering work of Teukolsky  \cite{Teukolsky,barpre}, done in the context of the Newman-Penrose formalism \cite{NP,chandra}, which partially received its mathematical foundation from Stewart and Walker  \cite{StewartWalker} and was subject to some extensions by other authors  \cite{Finley,GHP}.
Teukolsky found a separable master equation whose eigenfunction solutions essentially solve the problem of the massless perturbations of any spin for the Kerr black hole in terms of gauge- and tetrad-invariant quantities.  
For the spin 2 black hole case the connection between the master equation and the de Rham Laplacian of the Riemann tensor was first noted by Ryan \cite{Ryan}.
Here we introduce a master equation for the Kerr-Taub-NUT spacetime whose symmetries allow  
the separation of the equation into radial and angular parts, generalizing some previous results valid for the Kerr and Taub-NUT spacetimes, and use it to study the question of superradiant scattering modes.

\section{The Kerr-Taub-NUT metric}

The metric of the Kerr-Taub-NUT spacetime \cite{Miller} in Boyer-Lindquist-like coordinates 
$(x^0=t, x^1=r, x^2=\theta, x^3=\phi)$ is
\begin{eqnarray}
\label{metrica}
\rmd s^2
   &=&\frac{1}{\Sigma}(\Delta -a^2\sin^2\theta)\rmd t^2
    -\frac{2}{\Sigma}[\Delta A -a(\Sigma +a A)\sin^2\theta]\rmd t
\rmd \phi \nonumber \\ 
    &&-\frac{1}{\Sigma}[(\Sigma +a A)^2\sin^2\theta-A^2\Delta] \rmd \phi^2
    -\frac{\Sigma}{\Delta}\rmd r^2 -\Sigma d\theta^2
\ .
\end{eqnarray}
Here  $\Sigma$, $\Delta$ and $A$ are defined by

\begin{equation}
\Sigma = r^2 +(\ell +a \cos \theta)^2,\ 
\Delta = r^2-2Mr-\ell^2 + a^2,\ 
A = a \sin^2\theta -2\ell\cos\theta \ .
\end{equation}

Units are chosen such that $G=c=1$, so that $(M,a,\ell)$ all have the
same dimension of length.
The source of the gravitational field has mass $M$, angular momentum $J=Ma$
(i.e., gravitomagnetic dipole moment) along the $z$-direction, 
and gravitomagnetic monopole moment
$-\ell $.

The two solutions $r_\pm =M \pm \sqrt{M^2-a^2+\ell^2}$ 
of the equation $\Delta =0$ define the radii of the inner ($r_-$) and outer ($r_+$) horizons
when $ a^2 \le M^2 + \ell^2$. Our attention will be confined to the region outside the outer horizon: $r\ge r_+$. 
Adopting the Misner periodicity condition discussed by Miller \cite{Miller}, the time coordinate is assumed to be periodic with period $8\pi \ell$, while the angular coordinates have their usual ranges: $0\leq \theta \leq \pi$, $0\leq \phi <2\pi$.

The metric is of Petrov type D and a Kinnersley-like null frame \cite{kinnersley} 
\begin{eqnarray}
\label{frame}
l&=& \frac{1}{\Delta} [(\Sigma+aA) \partial_t+ \Delta\partial_r +a\partial_\phi]\ , \nonumber \\
n&=& \frac{1}{2\Sigma} [(\Sigma+aA) \partial_t -\Delta\partial_r +a\partial_\phi ] \ ,  \nonumber \\
m&=& \frac{1}{\sqrt{2}(\ell-ir+a\cos\theta)} [A\csc \theta \partial_t-i\partial_\theta+\csc \theta\partial_\phi ]
\end{eqnarray}
can be introduced to define Newman-Penrose (NP) quantities.
The only nonvanishing Weyl scalar is 
\begin{equation}
\label{psi2}
\psi_2=(M-i\ell)\rho^3
\end{equation}
and the only nonvanishing spin coefficients are
\begin{eqnarray}
\label{spincoeff}
\rho&=& -(r-i\ell-ia\cos\theta)^{-1}\ ,\quad \beta=-\rho^*\cot\theta/(2\sqrt 2)\ , 
\nonumber\\
\mu&=&\rho^2\rho^*\Delta/2\ ,\quad  \pi=ia\rho^2\sin\theta/\sqrt{2}\ ,
\nonumber\\
\alpha&=& \pi-\beta^* \ ,\quad  \tau=-ia\rho \rho^* \sin \theta /\sqrt{2}\ ,
\nonumber\\
\gamma&=&\mu+\rho\rho^* (r-M)/2 \ .
\end{eqnarray}

A master equation for the gauge- and tetrad-invariant first-order massless perturbations of any spin in this background can be given starting from the following Newman-Penrose relations for any vacuum type D geometry (here considered with no backreaction) \cite{Teukolsky} 
\begin{eqnarray}
\fl\quad\label{mia1}
&&
\{[D-\rho^{*}+\epsilon^*+\epsilon-2s(\rho +\epsilon)](\Delta+\mu-2 s \gamma)
\\ \fl\quad
&&
  -[\delta+\pi^{*}-\alpha^{*}+\beta
-2 s(\tau+\beta)] \,(\delta^{*}+\pi-2 s\alpha)
  -2(s-1)(s-1/2)\psi_{2}\}\Psi=0 
\nonumber
\end{eqnarray}
for spin weights $s=1/2,1,2$  and
\begin{eqnarray}\label{mia2}
\fl\quad
&&
\{[\Delta-\gamma^{*}+\mu^{*}-\gamma-2 s (\gamma+\mu)](D-\rho-2s\epsilon)
\\ \fl\quad
&&
-[\delta^{*}-\tau^{*}+\beta^{*}-\alpha
-2 s (\alpha+\pi)](\delta-\tau-2 s \beta)
-2(s+1)(s+1/2)\psi_{2}\}\Psi=0
\nonumber
\end{eqnarray}
for $s=-1/2,-1,-2$. The case $s=\pm 3/2$ can be derived instead  by following the work of 
G\"uven \cite{guven}, which is expressed in the alternative Geroch-Held-Penrose formalism \cite{GHP}. Finally the case $s=0$ is given by
\begin{eqnarray}\fl\quad
\meqalign{
&[D\Delta+\Delta D-\delta^* \delta-\delta\delta^*
+(-\gamma-\gamma^*+\mu+\mu^*)D+(\epsilon+\epsilon^*-\rho^*-\rho)\Delta \cr
&+(-\beta^*-\pi+\alpha+\tau^*)\delta+(-\pi^*+\tau-\beta+\alpha^*)\delta^*]\Psi=0\ .
\cr}
\end{eqnarray}
Note that only in these NP equations has the standard notation for the directional derivatives
$D=l^\mu \partial_\mu$, $\Delta=n^\mu \partial_\mu$ and  $\delta=m^\mu \partial_\mu$ been used and the second of these should not be confused  with the equally standard notation for the metric quantity $\Delta= r^2-2Mr-\ell^2 +a^2$ used everywhere else in this article.

As in the case of the Kerr spacetime \cite{bcjr}, representing a gravitomagnetic dipole, and the Taub-NUT spacetime \cite{TeukTN}, representing a gravitomagnetic monopole, all these equations for distinct spin weights can be cast into a single compact form in the 
Kerr-Taub-NUT spacetime as well, by introducing a ``connection vector" with components
\begin{eqnarray}
\fl\quad
&\Gamma^t =& \frac{1}{\Sigma}\left[\frac{(M+r)a^2+(M-3r)\ell^2+r^2(r-3M)}{\Delta }
+i\frac{2\ell+a\sin^2\theta\cos\theta}{\sin^2\theta}  \right]\ ,
\nonumber\\
\fl\quad
&\Gamma^r =& -\frac{1}{\Sigma}\,(r-M)\ ,
\nonumber\\
\fl\quad
&\Gamma^\theta =& 0\ , 
\nonumber\\
\fl\quad
&\Gamma^\phi =&-\frac{1}{\Sigma}\left[\frac{a(r-M)}{\Delta}+i\frac{\cos\theta}{\sin^2\theta }
\right]
\label{eq:SPINNOL}
\end{eqnarray}
satisfying
\begin{equation}
\nabla^\mu \Gamma_\mu = -\frac{1}{\Sigma}\ ,\quad
\Gamma^\mu \Gamma_\mu = \frac{\cot^2\theta}{\Sigma} +4\psi_2 
\label{eq:GAMMAPROP} \ .
\end{equation}
The resulting master equation has the form
\begin{equation}
\fl\qquad
[(\nabla^\mu+s\Gamma^\mu)(\nabla_\mu+s\Gamma_\mu)-4s^2\psi_2]\Psi=0\ ,\qquad 
{\textstyle s=0,\pm\frac12,\pm1,\pm\frac32,\pm2}\ , 
\label{eq:bellak}
\end{equation}
where $\psi_2$ is the Kerr-Taub-NUT background Weyl scalar given by (\ref{psi2}).
This master equation characterizes the common behavior of all these massless fields in this background differing only in the value of the spin-weight parameter $s$. In fact, the first term on its left-hand side represents (formally) a d'Alembertian, corrected by taking the spin-weight into account, and the second term is a (Weyl) curvature term also linked to the spin-weight value. 
Table \ref{TAB} shows the various Newman-Penrose quantities for which the master equation holds following the standard notation \cite{Teukolsky}, where in the spin-2 case $\psi_0$ and $\psi_4$ refer to the perturbed Weyl scalars.

\begin{table}
\begin{center}
\begin{tabular}{c||ccccccccc}
$s$ & 0 & 1/2 & -1/2 & 1 & -1 & 3/2 & -3/2 & 2 & -2 \\
$\Psi$ & $\Phi$ & $\chi_0$ & $\rho^{-1}\chi_1$ & $\phi_0 $ & $\rho^{-2}\phi_2$ &
$\Omega_0$ & $\rho^{-3}\Omega_3$ & $\psi_0$ & $\rho^{-4}\psi_4$ 
\end{tabular}
\end{center}
\caption{The spin-weight $s$ and the physical field component $\Psi$ for the master equation.}\label{TAB}
\end{table}

The standard notation of the Newman-Penrose formalism that is employed here will be confined to this section. Some of the symbols of this formalism will be used to designate different physical quantities in what follows. For example, $m$  and $\gamma$, used above in equations (\ref{frame}) and (\ref{spincoeff}),  will denote respectively an angular momentum parameter and the Lorentz gamma factor in the rest of this paper.

\section{Solution of the master equation}

Remarkably the master equation (\ref{eq:bellak}) admits separable solutions of the form
\begin{equation}
\psi(t,r,\theta,\phi)=e^{-i\omega t}e^{i m \phi} R(r)Y(\theta)\ ,
\end{equation}
where $\omega>0$ is the wave frequency and $m$ is the azimuthal separation constant (and eigenvalue of the usual angular momentum operator associated with the axial symmetry).
Since the $\phi$ coordinate has period $2\pi$, $m$ must be an integer in order for these solutions to be smooth. Assuming the Misner condition that the $t$ coordinate be periodic with period $8\pi \ell$, one is similarly led to the ``quantization condition" that $4\omega \ell$ also be an integer \cite{TeukTN}. 

The radial equation is then
\begin{equation}
\label{radeq}
\Delta^{-s} \frac{\rmd}{\rmd r}\left(\Delta^{s+1}\frac{\rmd R(r)}{\rmd r} \right)
+V_{\rm (rad)}(r)R(r)=0\ ,
\end{equation}
with
\begin{eqnarray}
\fl V_{\rm (rad)}(r) &=& s(s+1)+\Omega +\omega^2 (r^2+2Mr+7\ell^2)+\frac{[am-2\omega(Mr+\ell^2)]^2}{\Delta}\nonumber \\
\fl && +2is[\omega (r-M)+\frac{am(r-M)+2\omega(Ma^2-r(M^2+\ell^2))}{\Delta}] \ ,
\end{eqnarray}
where $\Omega$ is the separation constant. 
Clearly the solution $R(r)$ of this equation depends on the value of the spin weight $s$, so when convenient this dependence will be made explicit using the notation $R(r)\equiv R_s(r)$.
The radial potential can also be given a more compact form
\begin{eqnarray}
V_{\rm (rad)}(r) &=& a^2\omega^2-\lambda + \omega^2 P^2 +2is \omega \sqrt{\Delta} \frac{\rmd P}{\rmd r} \ ,
\end{eqnarray}
where $\lambda =-\Omega+a^2\omega^2-s(s+1)-4\omega^2\ell^2-2\omega a m$ and 
$P=[r^2+a^2+\ell^2-am/\omega]/\sqrt{\Delta}$.

Equation (\ref{radeq}) will be studied on the interval $r\in (r_+ , \infty )$, where the metric and the chosen tetrad (\ref{frame}) are well behaved, closely following the usual treatment of black hole perturbations that motivates the present investigation. Of course, our analysis can be extended to the whole spacetime, but this would require introducing new coordinates and tetrads and would further complicate the discussion.

By introducing the scaling
\begin{equation}
R(r)=(r^2+a^2+\ell^2)^{-\frac 12}\Delta^{-\frac s2}H(r)\equiv \mathcal{Q}_s^{-1}\, H(r)
\end{equation}
and the ``tortoise" coordinate transformation $r\to r_*$, where
\begin{equation}
\frac{\rmd r}{\rmd r_*}=\frac{\Delta}{r^2+a^2+\ell^2}\ ,
\end{equation}
the radial equation can be transformed into the one-dimensional Schr\"odinger-like equation
\begin{equation}
\label{newradi}
\frac{\rmd ^2 }{\rmd r_*^2}H(r)+\tilde V H(r)=0
\end{equation}
with the potential
\begin{eqnarray}
\fl
\tilde V&=&
\left[ \frac{K^2}{(r^2+a^2+\ell^2)^2} -G^2-\frac{\rmd G}{\rmd r_*}
         -\frac{2iKs(r-M)}{(r^2+a^2+\ell^2)^2}\right]
\nonumber \\
\fl
&& +\frac{\Delta}{(r^2+a^2+\ell^2)^2}
   \left[-\lambda + 4i\omega  rs \right]
\ ,
\end{eqnarray}
where
\begin{equation}
G=\frac{s(r-M)}{(r^2+a^2+\ell^2)}+\frac{r\Delta}{(r^2+a^2+\ell^2)^2}
=\frac{\rmd}{\rmd r_*} \ln \mathcal{Q}_s
\end{equation}
and $K=(r^2+a^2+\ell^2)\omega-am$
have been introduced in analogy with Teukolsky's treatment of the perturbations of the exterior Kerr spacetime.
It is useful to rewrite this potential in the following more compact form
\begin{eqnarray}
\fl\quad
\tilde V&=&
-\mathcal{Q}_s^{-1}\frac{\rmd^2}{\rmd r_*^2}\mathcal{Q}_s
+\frac{[K^2-2iKs(r-M) +\Delta (-\lambda +4i\omega  rs)]}{(r^2+a^2+\ell^2)^2}\ .
\end{eqnarray}

The asymptotic form of the radial equation as $r\to\infty \, (r_{*}\to\infty)$ is
\begin{equation}
\frac{\rmd ^2 }{\rmd r_*^2}H(r)+\left(\omega^2+\frac{2i\omega s}{r} \right ) H(r)=0\ ,
\end{equation}
which has asymptotic solutions $H\sim r^{\pm s}e^{\mp i\omega r_*}$, i.e. $R\sim e^{-i\omega r_*}/r$ and $R\sim e^{i\omega r_*}/r^{2s+1}$ (in accordance with the peeling theorem \cite{NP,Kroon}) and in this regime the effect of $\ell$ appears negligible.
On the other hand  close to the horizon $r\to r_+\, (r_{*}\to-\infty)$, the  asymptotic form of the radial equation becomes
\begin{equation}
\frac{\rmd ^2 }{\rmd r_*^2}H(r)+\left( k -i b_{+} \right )^2 H(r)=0\ ,
\end{equation}
where $b_{+}=\frac{s(r_+-M)}{2(M r_+ +\ell^2)}$, $k=\omega -m\omega_+$ and
\begin{equation}\label{omegahor} 
\omega_+=\frac{a}{r_+^2+a^2+\ell^2}=\frac{a}{2(M r_+ +\ell^2)}
\end{equation}
is the ``effective angular velocity" of the horizon.
The asymptotic solutions are $H\sim e^{\pm i(k -i b_{+})r_*}\sim \Delta^{\pm s/2}e^{\pm i k r_{*}}$, i.e. $R\sim e^{ik r_*}$ and $R\sim \Delta^{-s} e^{-ik r_*}$.
Only one of these two behaviors for $R$ is correct in the sense that it implies regularity of the fields on the horizon $r=r_+$ as stated by Teukolsky.
The asymptotic boundary conditions on the horizon are a delicate problem because both the coordinates and the tetrad being used in this discussion are singular there, but it was solved by Teukolsky in the Kerr case by picking out the second solution as having the correct behavior in that limit, which must therefore also be the correct choice in the present more general case. This result follows from the requirement of causality that at the horizon we must choose the boundary condition that the wave is always ingoing.

An immediate consequence of the behavior of the solution at the horizon and at spatial infinity is that there are superradiant scattered modes as in the Kerr case \cite{Teukolsky,Wald}, but now influenced by the nonzero value of the parameter $\ell$. For superradiant modes, reflected waves carry away more energy than the incident waves bring in, with the rotation of the black hole supplying the extra energy.

To understand how this comes about, consider first the $s=0$ case and imagine a solution of equation (\ref{newradi}) corresponding to the reflection and transmission of a radially incident wave of frequency $\omega >0$. For $r_*\to \infty$ one has 
\begin{equation}
\Phi \sim \mathcal{A}\, e^{-i\omega (t+r_*)}+\mathcal{A}\, \mathcal{R}\, e^{-i\omega (t-r_*)},
\end{equation}
where $\mathcal{A}$ is the amplitude of the incident wave and $\mathcal{R}$ is the reflection amplitude, 
while for $r_*\to -\infty$ one has
\begin{equation}
\label{transmitted}
\Phi\sim \mathcal{A}\, \mathcal{T}\, e^{-i\omega t-ik r_*},
\end{equation} 
where $\mathcal{T}$ is the transmission amplitude,
and the angular coordinates in $\Phi$ are suppressed.
The potential $\tilde V$ is real in this case ($s=0$); therefore, it follows from flux conservation that
\begin{equation}
|\mathcal{R}|^2+\frac{k}{\omega}|\mathcal{T}|^2=1.
\end{equation}
If $k/\omega <0$, then $|\mathcal{R}|>1$ and so one has superradiance.
The superradiance condition 
\begin{equation}
\label{sr}
k=\omega- m \omega_+ <0
\end{equation}
thus depends on the value $\omega_+$ of the effective angular velocity of the horizon. 
For any frequency $\omega>0$, this condition can be satisfied for large enough values of the azimuthal separation constant $m$ of the same sign as $\omega_+$ and $a$, i.e., when the angular momentum of the wave is in the same sense as the angular velocity of the horizon.
It follows from Eq.~(\ref{omegahor}) that, for fixed $a=J/M$, $r_+$ increases and hence $\omega_+$ decreases if $M$ or $|\ell|$ are increased. Thus for a spherical system where $a=0$ implies $\omega_+=0$, there is no superradiance, and in the KTN case where $a\neq0$, this phenomenon can also be suppressed by 
increasing $M$ and/or $|\ell |$ such that $\omega_+$ becomes so small that, for fixed $\omega$ and $m$, the quantity $k$  becomes positive.
In fact superradiance can be suppressed independently of the values of $M$ and $J$ by making $|\ell |$ sufficiently large. Note that the transmitted wave in Eq.~(\ref{transmitted}) moves toward the horizon 
for $k/\omega > 0$, but reverses direction in the case of superradiance with $k/\omega < 0$; moreover, total reflection occurs in the exceptional case that $k/\omega = 0$.

When the spin-weight $s$ is an integer, we would expect that there would still be superradiance because in the $\ell \to 0$ limit one obtains the Kerr case which is well known to exhibit this phenomenon for integer-spin fields.
In the case of half-integer spin fields, the Kerr case does not exhibit superradiance, but when  $\ell\neq0$ this effect could in principle change, so the question must be re-examined which we do here using a field theoretic approach \cite{chandra,guven,unruh}.
In fact one condition for having superradiance is that the flux of particles across the (null) horizon in the forward time direction be negative
\begin{equation}\label{eq:nJ}
( \xi_\mu \, J_{|s|}^\mu ) \vert_{r=r_+} < 0\ ,
\end{equation}
where $\xi=\partial_t+\omega_+ \partial_\phi$ is the future-directed (null) normal to the horizon and $J_{|s|}^\mu $ is the conserved particle number current vector 
associated with the various fields of spin $|s|$.
This condition corresponds to a positive flux out of the horizon.

Alternatively one may consider the rate $(dN/dt)_{\rm in}$ at which particles are falling in through the horizon per unit time, which must be negative for superradiance to occur 
\begin{equation}
\label{current}
\left(\frac{\rmd N}{\rmd t}\right)_{\rm in}
= -\int_{r_+} \sqrt{-g} J_{|s|}^r \rmd \theta \rmd \phi <0\ ,
\end{equation} 
where $g$ is the determinant of the spacetime metric. The appendix shows how this
integral relation follows from integrating the particle flux over a suitable region of the spacetime. 

For the scalar field case one has 
\begin{equation}
J_{0}{}_\mu = i \hbar (\Phi^* \nabla_\mu \Phi -  \Phi \nabla_\mu \Phi^* )\ ,
\end{equation}
so that
\begin{equation}\label{scalarflux}
(\xi_\mu \, J_{0}^\mu )\vert_{r=r_+}
= 2\hbar k\,|\mathcal{A}|^2 |\mathcal{T}|^2 |Y(\theta)|^2\ ,
\end{equation}
which is negative when  $k=\omega -m \omega_+<0$, corresponding to superradiance.
Analogously, evaluating the number of particles entering the horizon per unit time
gives
\begin{equation}
\label{scalcurr}
\left(\frac{\rmd N}{\rmd t}\right)_{\rm in}
= 2\hbar k \frac{a}{\omega_+}\, |\mathcal{A}|^2 |\mathcal{T}|^2 \ ,
\end{equation} 
which is negative under the same condition.
Note that
the angular part $Y(\theta)$ of the scalar field, which can be taken to be real, has been normalized by 
\begin{equation}
\int_0^\pi   Y^2(\theta)\,\sin \theta \,\rmd \theta =\frac{1}{2\pi} \ .
\end{equation}
In fact since $\Delta=0$ on the horizon, the surface element for a sphere with the horizon radius is just $(\Sigma+aA)|_{r=r_+} \rmd\theta \rmd\phi$, where
$(\Sigma+aA)|_{r=r_+} = r_+^2+a^2+\ell^2=  {a}/{\omega_+}$, so the integral of (\ref{scalarflux}) over the angular variables produces exactly (\ref{scalcurr}).

In contrast with the case of bosons, fermions do not exhibit superradiance, mirroring exactly the corresponding behavior in the Kerr spacetime. For the $|s|=1/2$ case, the neutrino particle number current is 
\begin{equation}
J_{1/2}{}^\mu =\bar \Psi \gamma^\mu \Psi\ ,
\end{equation}
where $\gamma^\mu $ are the (coordinate) Dirac matrices.
Passing to the more standard spinor formalism (see \cite{chandra} p.~539 for notation and conventions), this current becomes

\begin{equation}
\frac{1}{\sqrt{2}}\, J_{1/2}{}^{\mu}=\sigma^{\mu}_{AB'}(P^{A}\bar P^{B'}+Q^{A}\bar Q^{B'})
\end{equation}
with $Q^A=-P^A$ and $Q^{A'}=-\bar P^{A'}$ and
\begin{equation}
\sigma_{AB'}{}^\mu=
\frac{1}{\sqrt{2}}
\begin{pmatrix}
l^\mu  & m^\mu \cr
\bar m^\mu & n^\mu \cr 
\end{pmatrix} 
\end{equation}


In particular one finds
\begin{equation}
P^0=-\chi_1 \ , \qquad 
P^1=\chi_0  \ , \qquad 
\bar P^{0'}=-\chi_1^*  \ , \qquad 
\bar P^{1'}=\chi_0^*   \ .
\end{equation}
Since the spinor formalism can be immediately converted into the Newman-Penrose language, the neutrino current can be also be expressed as
\begin{equation}
\label{Current}
J_{1/2}{}^{\mu}=2[l^{\mu}\vert \chi_1\vert^2+n^{\mu}\vert \chi_0\vert^2-m^{\mu} (\chi_1\chi_0^*)-m^{*\mu}(\chi_0\chi_1^*)]\ .
\end{equation}
In this case ($|s|=1/2$) the condition for superradiance is never satisfied.
In fact following G\"uven \cite{guven} and using the relations
\begin{equation}
\vert \chi_0\vert^2 = |R_{1/2}(r)|^2 Y_{1/2}^2, \qquad \vert \chi_1\vert^2=\frac{1}{\Sigma}|R_{-1/2}(r)|^2 Y_{-1/2}^2,
\end{equation}
in (\ref{Current}) and (\ref{current}), one finds
\begin{equation}
\left(\frac{\rmd N}{\rmd t}\right)_{\rm in}
= \left( \Delta |R_{1/2}(r)|^2-2|R_{-1/2}(r)|^2 \right ) \, \vert_{r=r_+}\ .
\end{equation}
Next using our previous results for the behavior  of $R_{s}(r)$ near the horizon
\begin{equation}
|R_{1/2}(r)|^2\sim \frac{|c_{1/2}|^2}{\Delta},\qquad |R_{-1/2}(r)|^2\sim |c_{-1/2}|^2 \Delta
\end{equation}
with $c_{\pm 1/2}$ constants, one finds the final result
\begin{equation}
\left(\frac{\rmd N}{\rmd t}\right)_{\rm in}=  |c_{1/2}|^2 >0
\end{equation}
which means that there is no superradiance in this case.

In the Rarita-Schwinger case ($|s|=3/2$), the field is described by the  Majorana spinor-valued 1-form $\Psi^\alpha$ \cite{kam,blun} 
satisfying the  further conditions $\nabla_\alpha \Psi^\alpha=0$ and $\gamma_\alpha \Psi^\alpha=0$, and the current is 
\begin{equation}
J_{3/2}{}^\mu =\bar \Psi_\alpha \gamma^\mu \Psi^\alpha \ .
\end{equation}
We expect that a calculation similar to that valid
for the neutrino field would show that
\begin{equation}
\xi_\mu \bar \Psi_\alpha \gamma^\mu \Psi^\alpha\vert_{r=r_+}< 0\ , 
\end{equation}
namely that even in this case there is no superradiance.

Of course the study of the equations for the remaining (gauge- and tetrad-dependent) components of the various fields (i.e. $\phi_1$, $\psi_3$, etc.) and of theorems analogous to those found by Wald \cite{Wald}, Fackerell and Ipser \cite{FI}  and G\"uven \cite{guven} for the Kerr case  require some extensions here. 
For the Kerr spacetime these theorems essentially state that the master equation is enough to describe all of the relevant physics of the fields considered in the exterior spacetime. 

In the Kerr-Taub-NUT case the master equation completely describes the scalar ($s=0$) and neutrino ($s=\pm 1/2$) fields over all of the spacetime outside the horizon. For higher spin fields of course the master equation describes only the highest and lowest spin-weighted components, as specified in Table 1.
However, because of the ``peeling theorem" \cite{NP,PenroseRindler}, this equation is sufficient to understand the relevant physics of the fields of any spin considered at spatial infinity, in the sense that other spin-weighted components become negligible. Close to the horizon, one has instead only a subset of perturbations under control, namely those indicated in Table 1.

To complete our analysis of the perturbations on the Kerr-Taub-NUT spacetime, we need to discuss the angular equation
\begin{equation}
\label{angolare}
\frac{1}{\sin\theta}\frac{\rmd }{\rmd \theta} 
  \left(\sin\theta \frac{\rmd Y(\theta)}{\rmd \theta}\right)
  + V_{\rm (ang)}(\theta)Y(\theta)=0\ ,
\end{equation}
where
\begin{eqnarray}
\label{angolare1}
V_{\rm (ang)}(\theta)
&=&-\Omega-\frac {(s+2\ell\omega)2m\cos \theta  
+(s+2\ell\omega)^{2}+{m}^{2}}{\sin^2\theta}\nonumber \\
&& +a^2\omega^2\cos^2\theta +2a\omega (-s+2\ell\omega)\cos\theta\ \,.
\end{eqnarray}
This equation generalizes the spin-weighted spheroidal harmonics of Teukolsky\cite{Teukolsky,FC,novello}.
Here instead of the
parameter $s$ in the spin-weighted spheroidal harmonics, the combinations $s + 2 \ell \omega$ and
$-s + 2 \ell \omega$ appear.
In fact,  introducing a new variable 
$x=1+\cos\theta$ and the rescaling 
\begin{equation}
Y=x^{|m-s-2\omega\ell |/2}(2-x)^{|m+s+2\omega\ell |/2}X(x)\ ,
\end{equation}  
one gets
\begin{equation}
x(x-2)X''+(B_1x+B_0)X'+(C_2x^2+C_1x+C_0)X=0\ ,
\end{equation}
where
\begin{eqnarray}
\fl
B_0&=& -2(|m-s-2\omega\ell|+1)\ ,\nonumber \\
\fl
B_1&=& 2+|m-s-2\omega\ell|+|m+s+2\omega\ell|\ ,\nonumber \\
\fl
C_0&=& \frac12 \,[m^2+(s+2\omega\ell)^2+|m^2-(s+2\omega\ell)^2|
+|m+(s+2\omega\ell)|+|m-(s+2\omega\ell)|]\ ,\nonumber \\
\fl &&+\Omega -a\omega [a\omega+2(s-2\omega\ell)]\ ,\nonumber \\
\fl
C_1&=& 2a\omega (a\omega +s -2\omega\ell)\ ,\nonumber \\
\fl
C_2&=&-a^2\omega^2\ ,
\end{eqnarray}
which is a generalized spheroidal wave equation of the Leaver form \cite{leav}.

\section{Gravitomagnetism}

We note that these results contain the combinations $\pm s\ + 2\omega \ell$, so that, in a certain sense, the spin is coupled to the gravitomagnetic monopole moment; for $a=0$ only the combination $s+2\omega \ell$ is involved. This is a novel manifestation of the spin-gravity coupling \cite{mas74,mas00}. 

To interpret this coupling let us start with the propagation of test electromagnetic fields in the KTN spacetime. Using the Skrotskii formalism \cite{skro} in which the gravitational field may be replaced by an equivalent material medium, the problem reduces to Maxwell's equations in a certain gyrotropic medium in a global inertial frame with Cartesian coordinates $(t,x,y,z)$. 
In this section only we employ the other metric signature $-+++$, which is more standard for the following considerations.
To simplify matters we assume that the electromagnetic waves have a time dependence of the form $e^{-i\omega t}$; moreover, we linearize the KTN metric. With these simplifications, the wave equation becomes
\begin{equation}
\label{dirac}
[\mathbf{P}+2\hbar \omega \mathbf{A}_{\rm (g)}]\times \mathbf{W}^\pm=\mp i\hbar \omega \mathcal{N} \mathbf{W}^\pm\ ,
\end{equation}
where $\mathbf{P}=-i\hbar \bfnabla $, $\mathbf{W}^\pm =\mathbf{E}\pm i \mathbf{H}$,
\begin{equation}
\mathbf{A}_{\rm (g)} 
= \frac{\mathbf{J}\times \bfrho }{|\bfrho |^3} 
+\ell{}z 
\frac{\hat{\mathbf{z}}\times \hat{\bfrho}}{x^2+y^2}\ , 
\end{equation}
and $\mathcal{N}\simeq 1+\frac{2M}{|\bfrho |}$ is the index of refraction of the medium. Here $\mathbf{E}$ and $\mathbf{H}$
are complex fields and $|\bfrho |=\sqrt{x^2+y^2+z^2}$ is the isotropic radial coordinate in the linearized KTN spacetime.
Furthermore, $\mathbf{W}^+$ represents the positive helicity wave amplitude and $\mathbf{W}^-$ represents the negative helicity wave amplitude. It is a general result that helicity is conserved in pure gravitational scattering.
A more detailed treatment is available elsewhere \cite{bcjmKTN}.

As stated above, a satisfactory interpretation of the KTN spacetime requires that the time coordinate $t$ be periodic with period $8\pi \ell$. Thus a propagating wave with a time dependence of the form $e^{-i\omega t}$ is possible only when $4\omega \ell $ is an integer. Eq.~(\ref{dirac}) can be expressed in the Dirac form and then interpreted as the wave equation for a particle of inertial mass $m_0=\hbar \omega$ and gravitomagnetic charge $q_0=-2\hbar \omega$ propagating in a gravitomagnetic field $\mathbf{B}_{\rm (g)}= \bfnabla \times \mathbf{A}_{\rm (g)}$ given by 
\begin{equation}
\mathbf{B}_{\rm (g)}
= \frac{J}{|\bfrho |^3}[3 (\hat{\bfrho} \cdot \hat{\mathbf{J}})\hat{\bfrho} -\hat{\mathbf{J}}]
-\ell \frac{\hat{\bfrho}}{|\bfrho |^2} \ .
\end{equation}	

Restricting our attention now to the motion in a purely monopole field (i.e. $J=0$) of strength $\mu_0=-\ell $, we note that the {\it classical} equation of motion would be
\begin{equation}
\dot{\mathbf{p}} = q_0 \mu_0 \frac{\bfrho \times \mathbf{v}}{|\bfrho |^3}\ ,
\end{equation} 
where $\mathbf{p}=m_0 \gamma \mathbf{v}$ is the kinetic momentum. It is well known that this equation has a constant of the motion given by 
$\mathbfcalJ = \bfrho \times \mathbf{p}-q_0\mu_0 \hat{\bfrho}$, 
which can be interpreted as implying that the orbital angular momentum of the particle is augmented, through its interaction with the monopole, by 
$\mathbf{S}'=-q_0 \mu_0 \hat{\bfrho} =2 \hbar \omega \ell \hat{\bfrho}$. Note that the magnitude of this vector $S'=|\mathbf{S}'|$ is a positive integer multiple of $\hbar/2$, since $4\omega \ell$ is an integer.
The intrinsic rotational symmetry of the monopole thus leads to a conservation law for the total angular momentum of the particle $\mathbfcalJ$ that consists of an orbital part plus a contribution from the angular momentum of the total field (generated by the monopole and the particle) that is acquired by the particle through its interaction with the monopole.
That this mechanical result extends to the wave treatment of the scattering problem has been demonstrated by a number of authors in the case of magnetic monopoles (see e.g. \cite{asg65, lb-nz98}). 

Note that the canonical momentum in the present case would be
$\mathbf{P}=\mathbf{p}+q_0 \mathbf{A}_{\rm (g)}=\mathbf{p}-2 \hbar \omega \mathbf{A}_{\rm(g)}$. The total angular momentum for a spinning quantum particle turns out to be 
\begin{equation}
\label{jtot}
\mathbfcalJ=\bfrho \times [\mathbf{P}+2 \hbar \omega \mathbf{A}_{\rm(g)}]+\mathbf{S}+\mathbf{S}'=\mathbf{L}+\mathbf{L'}+\mathbf{S}+\mathbf{S}', 
\end{equation}
where $\mathbf{L}=\bfrho \times \mathbf{P}$ (satisfying  $[L_i,L_j]=i\hbar \,\epsilon_{ijk}\, L_k$) is the orbital part, $\mathbf{S}$ is the spin part,  
$\mathbf{L'}=2 \hbar \omega \bfrho \times \mathbf{A}_{\rm(g)}=-2\hbar \omega \ell \cot \theta \, \hat \bftheta$ and $\mathbfcalJ$ (satisfying $[\mathcal{J}_i,\mathcal{J}_j]=i\hbar \, \epsilon_{ijk}\, \mathcal{J}_k$) is the generator of spatial rotations. 
Heuristically one may say that the spinning particle, in its interaction with the monopole, picks up an additional spin contribution $\mathbf{S}'=2 \hbar \omega \ell \hat{\bfrho}$, such that the net effective spin of the particle along any radial direction (like the $z-$ axis)---taken to be the axis of quantization---would be $S+S',\ldots , -S+S'$.
The general situation is, however, not so straightforward since $\mathbfcalJ=\mathbf{L}+\mathbf{L'}+\mathbf{S}+\mathbf{S}'$
and we note that 
\begin{equation}
[L_i+L'_i,L_j+L'_j]=i\hbar \epsilon_{ijk}(L_k+L'_k+S'_k)\ .
\end{equation}
Thus the separation of the total angular momentum (\ref{jtot}) into orbital and spin parts is not quite obvious in this case. Nevertheless, since the master equation describes the behavior of the highest and the lowest 
spin-weighted amplitudes, it is natural to expect that the angular part of the equation would only contain the spin-weight combinations $s + 2 \omega \ell$ and $-s + 2 \omega \ell$.

\section{Discussion}

A master equation for the gauge- and tetrad-invariant first-order massless perturbations of any spin $s\leq2$ on the Kerr-Taub-NUT background spacetime has been obtained and separated.
We have studied superradiance in this case and have shown that the situation is very similar to the Kerr spacetime; in particular, we have demonstrated the absence of superradiance for half-integer spin perturbations. Furthermore, the interaction of the perturbing field with the gravitomagnetic monopole contributes a certain half-integer spin component to the angular momentum that combines with the spin of the field; this novel form of spin-gravity coupling has been briefly discussed here. 
This investigation offers the possibility of achieving a better understanding of perturbations of black hole spacetimes within this larger family.
That is, one can  in principle extend the methods developed in \cite{futt} to discuss various aspects of scattering of radiation from the Kerr-Taub-NUT spacetime such as the polarization properties of the scattered radiation, glory effects and quasinormal mode oscillations. These may then be used to search for rotating gravitomagnetic monopoles.

The source of the Taub-NUT solution is a gravitational dyon (with gravitoelectromagnetic monopoles $m$ and   $-\ell$). The Taub-NUT solution can be extended to include a cosmological constant \cite{DemNew}
or an infinite set of multipole moments pertaining to axisymmetric deformations of a rotating source \cite{quemas}. The periodicity of the time coordinate in the Taub-NUT spacetime renders such sources, if they exist at all, rather exotic astrophysical systems. Normal astronomical systems do not exhibit such periodicity; therefore, the temporal periodicity may be associated with Hubble-scale structures \cite{zolinbel97} or compact dark-matter candidates \cite{rahnz}; in these connections, lensing properties of Taub-NUT spacetime have been studied in detail \cite{zolinbel97,rahnz} and compared in the latter case with astronomical observations \cite{rahnz}. 

A remark is in order here concerning an indirect application of Kerr-Taub-NUT spacetime to rotating relativistic disks, which are of great interest in astrophysics. The KTN spacetime can be employed to generate exact solutions of the Einstein-Maxwell equations corresponding to stationary axially symmetric disklike configurations of matter with a magnetic field \cite{lete, gonzlete}.

The Taub-NUT spacetime has had significant applications in theoretical studies of the spacetime  structure in general relativity and, more recently, in quantum gravity. Its Euclidean extension is important for the study of monopoles in gauge theories. Embedding the Taub-NUT gravitational instanton into five-dimensional Kaluza-Klein theory leads to a Kaluza-Klein monopole \cite{sor,groper,vamvis}. Euclidean Taub-NUT spaces have been discussed by a number of authors in connection with monopoles in supersymmetric gauge theories \cite{gibman}. Further generalizations and extensions of Taub-NUT spaces (such as the Kerr-Newman-Taub-NUT-adS spaces) are topics of current research in string theory \cite{string,sen,susy}.

\appendix

\section{Integral form of the particle number current conservation law}

This article has dealt exclusively with the exterior KTN spacetime. Therefore, in
the determination of the number of particles falling in through its horizon per unit time, one
needs to apply the conservation law for the particle number current only  in the exterior region. To
this end, consider the volume enclosed by two ``spheres" of time-independent radii $r_+\leq R_1 <R_2$.
Eq.~(\ref{current}) can be derived easily from the
integral of the particle number conservation equation 
\begin{equation}
  0 = \sqrt{-g} J^\alpha{}_{;\alpha}
    = \partial_\alpha (\sqrt{-g} J^\alpha)
    = \partial_t  (\sqrt{-g} J^t) + \partial_i  (\sqrt{-g} J^i)
\end{equation}
over the region $V$ within a time coordinate hypersurface
bounded by these two spheres, and then using Gauss's law to convert the second term to a surface integral over the boundary
\begin{eqnarray}
\frac{dN_{V}}{dt}
&\equiv&
\frac{\rmd}{\rmd t} \int_{V} \sqrt{-g} J^t\, \rmd r \rmd\theta \rmd\phi
  = -\int_{V} \partial_i  (\sqrt{-g} J^i) \rmd r \rmd\theta \rmd\phi \cr
&=& -\int_{r=R_2} \sqrt{-g} J^r \rmd\theta \rmd\phi
  +\int_{r=R_1} \sqrt{-g} J^r \rmd\theta \rmd\phi \ .
\end{eqnarray}
This gives the rate of change of the number $N_V$ of particles in this region in terms of the flux entering the outer sphere and exiting the inner sphere.

When the integration domain is replaced by  the  region enclosed between the horizon and a large sphere at infinity, one may compute the rate at which particles are leaving this region through the horizon alone. The contribution from the horizon, i.e.,
\begin{equation}
   \int_{r=r_+} \sqrt{-g} J^r \rmd\theta \rmd\phi
= \frac{dN_{V}}{dt}
 + \int_{r=\infty} \sqrt{-g} J^r \rmd\theta \rmd\phi\ ,
\end{equation}
when sign-reversed,
gives the rate at which particles are entering the horizon,
leading to Eq.~(\ref{current}).
This sign-reversed quantity must be negative for superradiance to occur.

\end{document}